%% file: INI_Globecom_v1.tex
\documentclass[journal]{IEEEtran}

\ifCLASSINFOpdf
   \usepackage[pdftex]{graphicx} 
\else
  \usepackage[dvips]{graphicx}
   declare the path(s) where your graphic files are
  \graphicspath{{../eps/}}
  
\fi

\usepackage[cmex10]{amsmath}
\usepackage{caption}
\usepackage{float}

\usepackage[table]{xcolor}
\usepackage{multirow}
\usepackage{multicol}
\usepackage{cuted}

\usepackage{enumerate}
\usepackage{subfigure}
\usepackage{acro}
\usepackage{epstopdf}
\usepackage{cite}
\usepackage{xcolor}
\usepackage{siunitx}

\usepackage{amsfonts}

\usepackage{gensymb}

\input{AcronymList}

\begin{document}

%
% paper title
\title{Inter-Numerology Interference Analysis for 5G and Beyond}

\author{
\IEEEauthorblockN{Abuu~B.~Kihero\IEEEauthorrefmark{1}, Muhammad~Sohaib~J.~Solaija\IEEEauthorrefmark{1}, Ahmet Yazar\IEEEauthorrefmark{1} and H\"{u}seyin Arslan\IEEEauthorrefmark{1}\IEEEauthorrefmark{2}}\\
\IEEEauthorblockA{\IEEEauthorrefmark{1}Department of Electrical and Electronics Engineering, Istanbul Medipol University, Istanbul, 34810 Turkey\\
}
\thanks{This work has been submitted to the IEEE for possible publication. Copyright may be transferred without notice, after which this version may no longer be accessible.}
\IEEEauthorblockA{\IEEEauthorrefmark{2}Department of Electrical Engineering, University of South Florida, Tampa, FL 33620 USA\\
Email: \{abkihero, msolaija\}@st.medipol.edu.tr,\{ayazar, huseyinarslan\}@medipol.edu.tr}

}

% make the title area
\maketitle

\begin{abstract}
One of the defining characteristics of 5G is the flexibility it offers for supporting different services and communication scenarios. For this purpose, usage of multiple numerologies has been proposed by the \ac{3GPP}. The flexibility provided by multi-numerology system comes at the cost of additional interference, known as inter-numerology interference (INI). This paper comprehensively explains the primary cause of INI, and then identifies and describes the factors affecting the amount of INI experienced by each numerology in the system. These factors include subcarrier spacing, number of used subcarriers, power offset, windowing operations and guard bands.
% This paper describes in a comprehensive manner how the interference is affected by factors such as subcarrier spacing, number of used subcarriers, power offset, windowing and guard bands.

\end{abstract}
	

\section{Introduction}
5G is expected to act as a platform enabling wireless connectivity to all kinds of services. The different service classes defined for 5G include eMBB (enhanced Mobile BroadBand), mMTC (massive Machine Type Communications) and URLLC (Ultra-Reliable and Low Latency Communications) \cite{3GPP_38_913}. These scenarios have their own specific demands causing 5G to have a wide range of requirements which dictates the need for a high degree of flexibility in the radio and network designs \cite{dahlman20164g}.  	  	

One of the steps towards achieving the required flexibility in 5G systems is the introduction of multi-numerology concept under the umbrella of 5G New Radio (5G-NR). The term numerology in 5G refers to a set of parameters like subcarrier spacing, symbol length and cyclic prefix in \ac{OFDM}. Table \ref{tab:numerology_structure} provides a summary of the properties of different NR numerologies as presented in \cite{3GPP_38_211} and \cite{3GPP_38_104}. Usage of multiple numerologies significantly affects the performance of the system. These effects include spectral efficiency, scheduling complexity, computational complexity, and signaling overhead \cite{yazar2018flexibility}. Employing multiple numerologies also introduces non-orthogonality into the system, causing interference between users belonging to different numerologies.

\begin{table}[h]
\centering
\caption{Numerology Structures for Data Channels in 5G\cite{3GPP_38_211}}
\label{tab:numerology_structure}
\begin{tabular}{|c|c|c|}
\hline
\begin{tabular}[c]{@{}c@{}}$\Delta f$\\ $(\si{\kilo\hertz})$\end{tabular} & \begin{tabular}[c]{@{}c@{}}$T^{CP}$\\ ($\si{\micro\second}$)\end{tabular} & \begin{tabular}[c]{@{}c@{}}Slot Duration\\ ($\si{\milli\second}$)\end{tabular} \\ \hline
15                                                                        & 4.76                                                                      & 1                                                                              \\ \hline
30                                                                        & 2.38                                                                      & 0.5                                                                            \\ \hline
60                                                                        & 1.19 $|$ 4.17                                                             & 0.25                                                                           \\ \hline
120                                                                       & 0.6                                                                       & 0.125                                                                          \\ \hline
\end{tabular}
\end{table}

Interference in multi-numerology systems, also called \ac{INI} has garnered increasingly more attention in recent times. An INI model is presented in \cite{zhang2018mixed} which describes INI as a function of frequency response of the interfering subcarrier; frequency offset between the interfering and the victim subcarriers, and the overlap in transmitter and receiver windows of the interferer and victim, respectively. Though detailed, this model is limited to windowed-OFDM system. Similarly, \cite{pekoz2017adaptive} uses adaptive windowing to minimize the interference and \cite{demir2017impact} tries to optimize the guard band and time keeping in view the power offset and requirements of the users. While these works have shed some light on the phenomenon of INI, a study which accounts for and individually explains all the factors contributing to INI is still lacking. Such a study is imperative as it would enable the development of efficient interference cancellation techniques for multi-numerology systems in 5G and beyond. In this paper, we attempt to address the above mentioned gap in the present literature by contributing the following:
\begin{itemize}
    \item An extensive discussion on synchronization and orthogonality issues of multi-numerology systems is provided.
    %\item A comprehensive explanation of non-orthogonality, which arises due to the use of multiple numerologies, is provided.
   \item The factors that affect INI are identified and their effects are explained in light of simulation results.
    \item This work also presents research opportunities regarding interference in multi-numerology systems. 
    %This work lays the foundation of a thorough mathematical model of INI in multi-numerlogy systems.
\end{itemize}
%This paper gives a concise description and explanation of the parameters that affect interference in multi-numerology systems.

The rest of the paper is organized as follows: Section \ref{sec:system_model} describes the system model used in this study and the assumptions that form its basis. Section \ref{sec:Synchronization} discusses the effects of multiple numerologies from orthogonality and synchronization perspective.
This is followed by highlighting the parameters that govern \ac{INI}, accompanied by simulation results and intuitive interpretation of each of them in Section \ref{sec:Factors}. Section \ref{sec:Conclusion} summarizes our findings and indicates the possible future direction of research.

\begin{figure*}
\centering
\includegraphics[width=2.0\columnwidth]{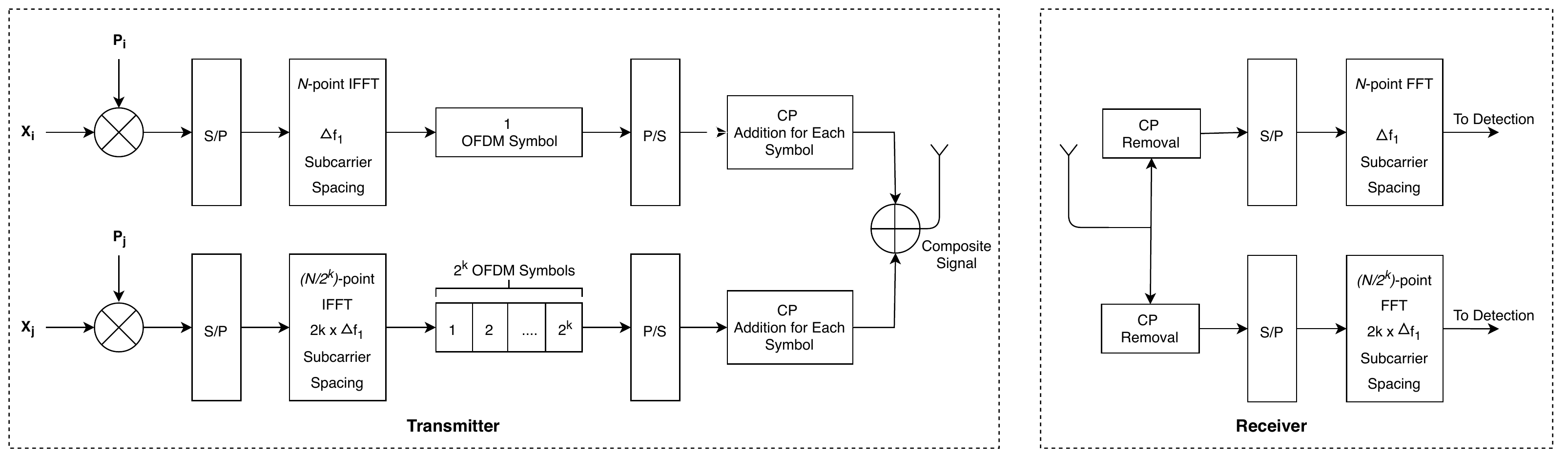}
\caption{Block Diagram of the Multi-numerology Implementation \cite{Yazar2018Flexible}}
\label{fig:block_diagram}
\end{figure*}

\section{System Model and Assumptions}
\label{sec:system_model}
Multi-numerology is a key concept of the 5G-NR frame structure. Our system model considers two numerologies which is the base case for multi-numerology systems. It can be generalized to any number of numerologies by considering one pair at a time. Each numerology block consists of multiple user equipments (UEs) which are non-overlapping in the frequency domain. It is assumed that the UEs have gone through a numerology selection process based upon the user and service requirements which may lead to different power levels amongst the users. This may be achieved by algorithms such as the one presented in \cite{yazar2018flexibility}.
In our model each numerology is assumed to cater to three users where the users of a particular numerology occupy equal bandwidth.
The data generated consists of binary phase shift keying (BPSK) symbols. Since choice of modulation scheme is not our primary concern for the time being, we have limited ourselves to BPSK because of its simplicity. Fig. \ref{fig:block_diagram} shows the block diagram of the multi-numerology implementation used in this paper \cite{Yazar2018Flexible}. $X_i$ and $X_j$ are complex modulated symbols for users $i$ and $j$ of numerology-1 and numerology-2, respectively. The users indices are defined as $i= 1, 2, ..., Q$ and $j = 1, 2, ..., R$, where $Q$ and $R$ are number of users scheduled in the corresponding numerologies. $P_i$ and $P_j$ are power ratios for $i^{th}$ and $j^{th}$ users in each numerology. 
The first numerology employs subcarrier spacing $\Delta f_1$ and N-point inverse Fourier transform (IFFT) while the second numerology's subcarrier spacing and IFFT size are scaled by a factor of $2^k$ and $1/2^k$ respectively, where $k$ is a positive integer. Similarly, the number of OFDM symbols for the second numerology is upscaled by $2^k$ as compared to first numerology. The IFFT operation is followed by addition of cyclic prefix (CP) with a certain ratio, $CP_R$, at the beginning of each symbol. 
In this study, we have narrowed down our focus on the factors affecting inter-numerology interference (INI) without considering noise or a wireless channel. The receiver removes CP before taking $N$-point and $N/2^k$-point FFT for first and second numerology, respectively. 
%and FFT of the received signal for the first numerology. For the second numerology, $N/2^k$ samples are used for FFT of size $N/2^k$. The composite signal is divided into $2^k$ parts called sub-blocks which are processed sequentially. %Fig. \ref{fig:block_diagram} shows the block diagram of the multi-numerology implementation used in this paper \cite{Yazar2018Flexible}.
We have employed Monte Carlo method to observe and analyze the interference statistics for each used subcarrier over 500 independent trials for each scenario discussed in following sections.

\section{Inter Numerology Synchronization and Orthogonality} \label{sec:Synchronization}
\begin{figure}[h]
	\centering
	\subfigure[Synchronization with individual CP]{
		\includegraphics[width=2.5in]{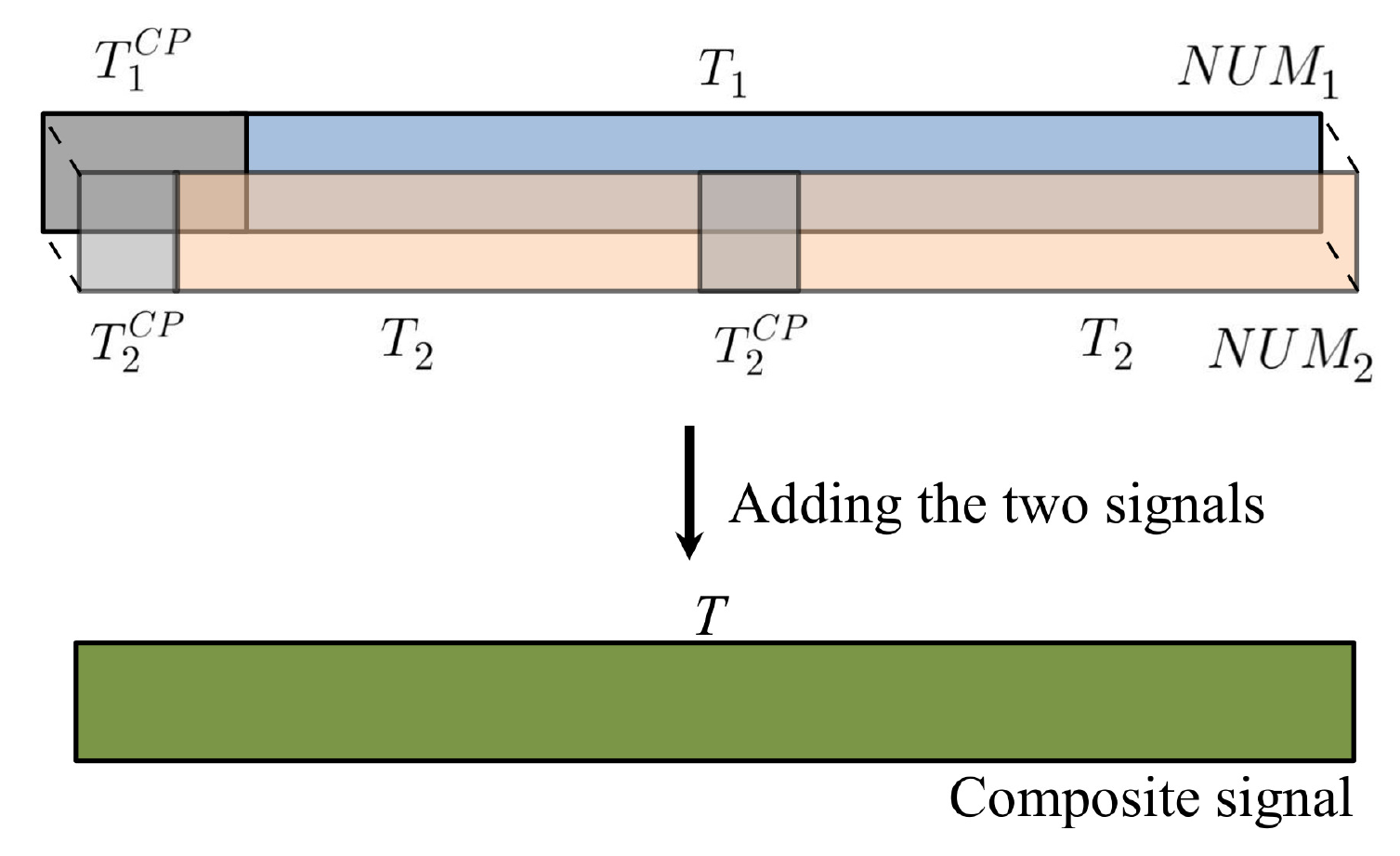} \label{fig:individualCP}
	}
	\subfigure[Synchronization with common CP]{
		\includegraphics[width=2.5in]{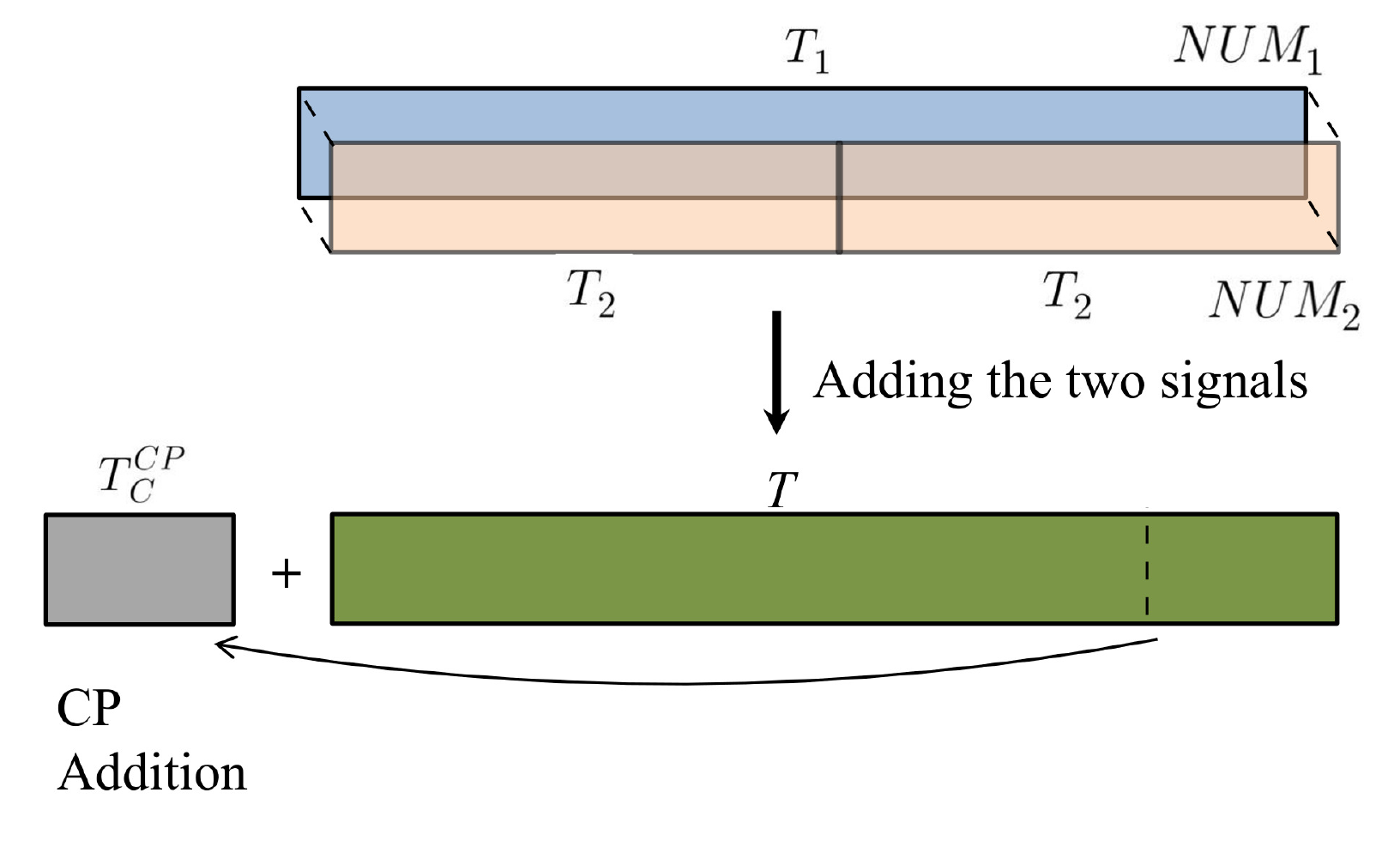} \label{fig:CommonCP}
	}
	\caption{Synchronizing Symbols of $NUM_1$ and $NUM_2$ for $k$ = 1}
	\label{fig:SymbolSync}
\end{figure}

Symbol lengths among numerologies tend to vary due to the usage of different \ac{ScS} which in turn makes the whole system  unsynchronized in time domain. Difficulty in achieving synchronization of OFDM symbols of different numerologies is one of the major drawbacks of  multi-numerology systems. However, if \ac{ScS} of one numerology is integral multiple of the \ac{ScS} of the other numerology, synchronization can be achieved over the so called \ac{LCM} symbol duration \cite{zhang2017subband}. For instance, if \ac{ScS} of numerology 1 ($NUM_1$), $\Delta f_1$, and that of numerology 2 ($NUM_2$), $\Delta f_2$, are such that $\Delta f_2$ $=$ $2^k\cdot\Delta f_1$, then, $2^k$ symbols of $NUM_2$ can be perfectly synchronized with 1 symbol of $NUM_1$, i.e $2^k\cdot T_2$ = $T_1$ where $T$ is the symbol duration. In this case $T_1$ is the \ac{LCM} symbol duration. Synchronization over \ac{LCM} symbol duration can be achieved in two ways:- \\
\textbf{\textit{By using individual CPs:}} This is the conventional way of creating synchronous composite signal of $NUM_1$ and $NUM_2$ in 5G where CPs are added, in accordance with $CP_R$, for all symbols of each numerology before the creation of the composite signal, as shown in Fig. \ref{fig:individualCP} (for $k = 1$). In this case, duration $T$ of the synchronous symbols can be written as

\begin{equation}\label{eq:individualCP}
T = T_1 + T^{CP}_{1} = 2^k\cdot (T_2 + T^{CP}_{2}).
\end{equation}
\textbf{\textit{By using common CP:}} This is another way of achieving synchronization over \ac{LCM} symbol duration. In this approach, \ac{MSE-OFDM} is adopted in $NUM_2$, which allows one CP to be used for $2^k$ OFDM symbols \cite{chouinard2005mse, nemati2018low}. CP size of $NUM_2$, $T^{'CP}_{2}$, is then determined from the resultant length of the concatenated $2^k$ symbols, which makes $T^{'CP}_{2}$ = $2^k\cdot T^{CP}_{2}$. In this case, a common CP can be used for both numerologies as shown in \cite{abusabah2018noma}. The common CP of length $T^{CP}_{c}$ = $T^{CP}_{1}$ = $T^{'CP}_{2}$, is appended after creation of the composite signal of $NUM_1$ and $NUM_2$ as show in Fig. \ref{fig:CommonCP}. The duration $T$ in this case is given by

\begin{equation}\label{eq:commonCP}
T = T_1 + T^{CP}_{c} = 2^k\cdot T_2 + T^{CP}_{c}.
\end{equation}
However, due to the adoption of \ac{MSE-OFDM} in $NUM_2$, an extra FFT and IFFT blocks are required at $NUM_2$ receiver to facilitate proper equalization and data detection as discussed in \cite{chouinard2005mse} and \cite{nemati2018low}.

\begin{figure}[h]
	\centering
	\includegraphics[width=3.in]{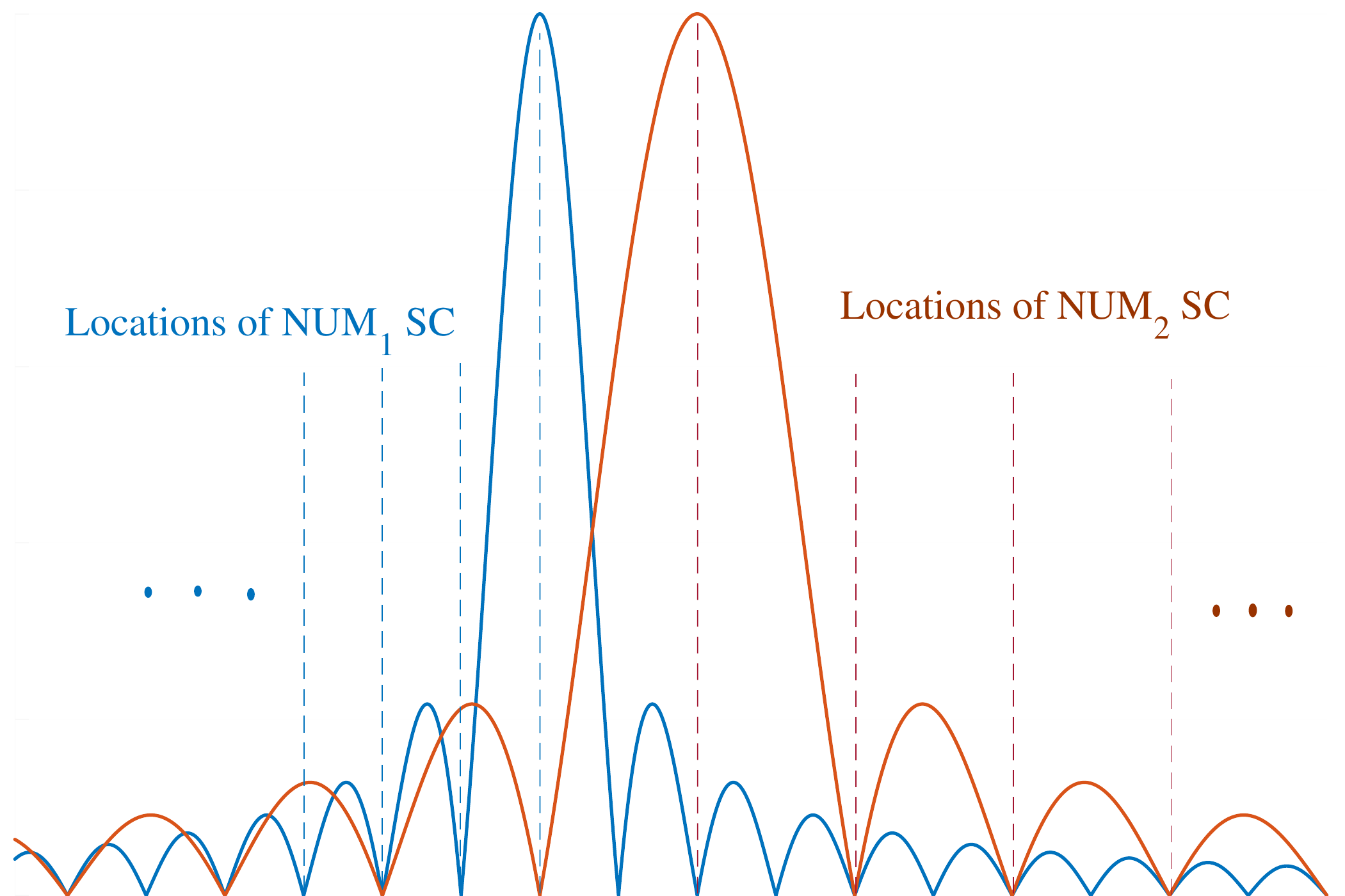}
	\caption{Multiplexed subcarriers of $NUM_1$ ($\Delta f_1$ = 15 kHz) and $NUM_2$ ($\Delta f_2$ = 30 kHz) at the transmitter (before composite signal) with guard band = $2*\Delta f_1$.}
	\label{fig:SCsTx}
\end{figure}

To understand how \ac{INI} affects multi-numerology signal synchronized by either of the above discussed approaches, let us first understand the coexistence of $NUM_1$ and $NUM_2$ subcarriers at the transmitter side before creation of the composite signal. We observe from Fig. \ref{fig:SCsTx} that $NUM_1$ causes no interference at any of the $NUM_2$ subcarriers, while $NUM_2$ imparts some interference on one out of every two subcarriers of $NUM_1$. Number of $NUM_1$ subcarriers affected by INI from $NUM_2$ depends on the ratio $\Delta f_1$/$\Delta f_2$ of the two numerologies.

\begin{figure}[h]
	\centering
	\subfigure[Common CP case]{
	    \includegraphics[width=2.2in]{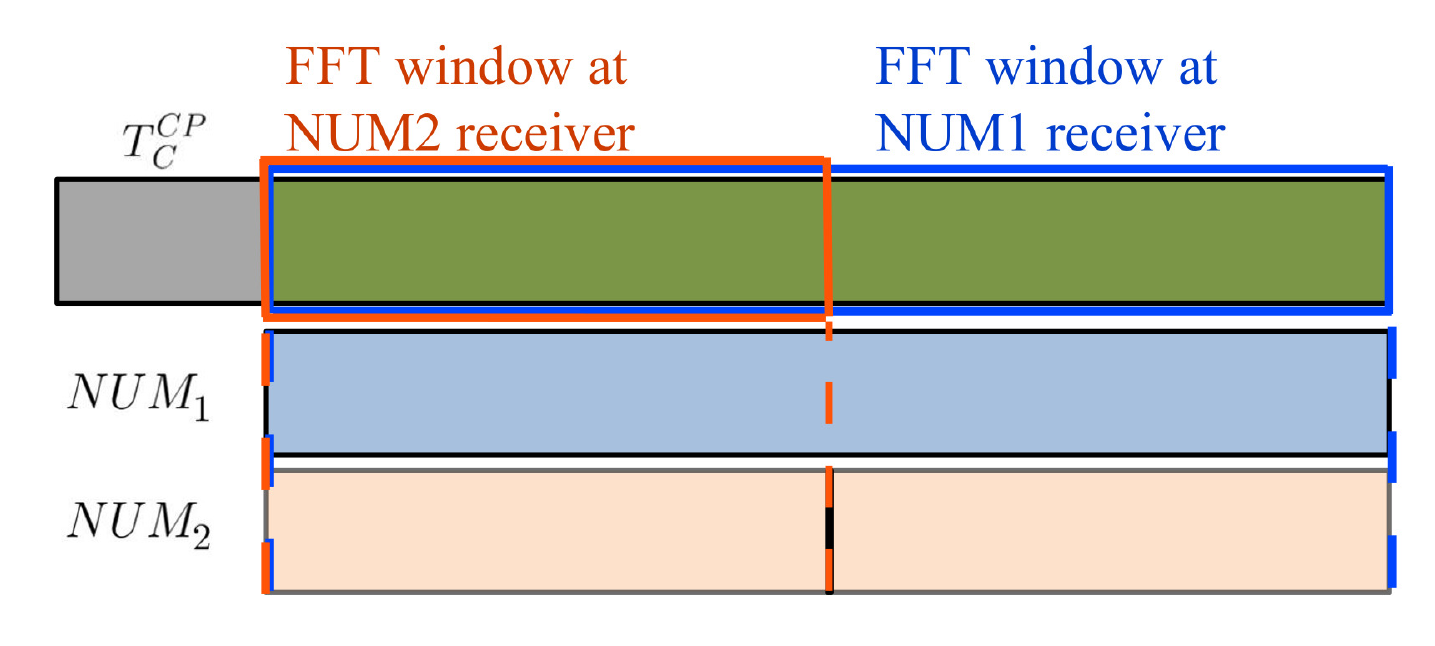} \label{fig:CommonCPreceiver}
	}
	\subfigure[Individual CP case]{
		\includegraphics[width=2.5in]{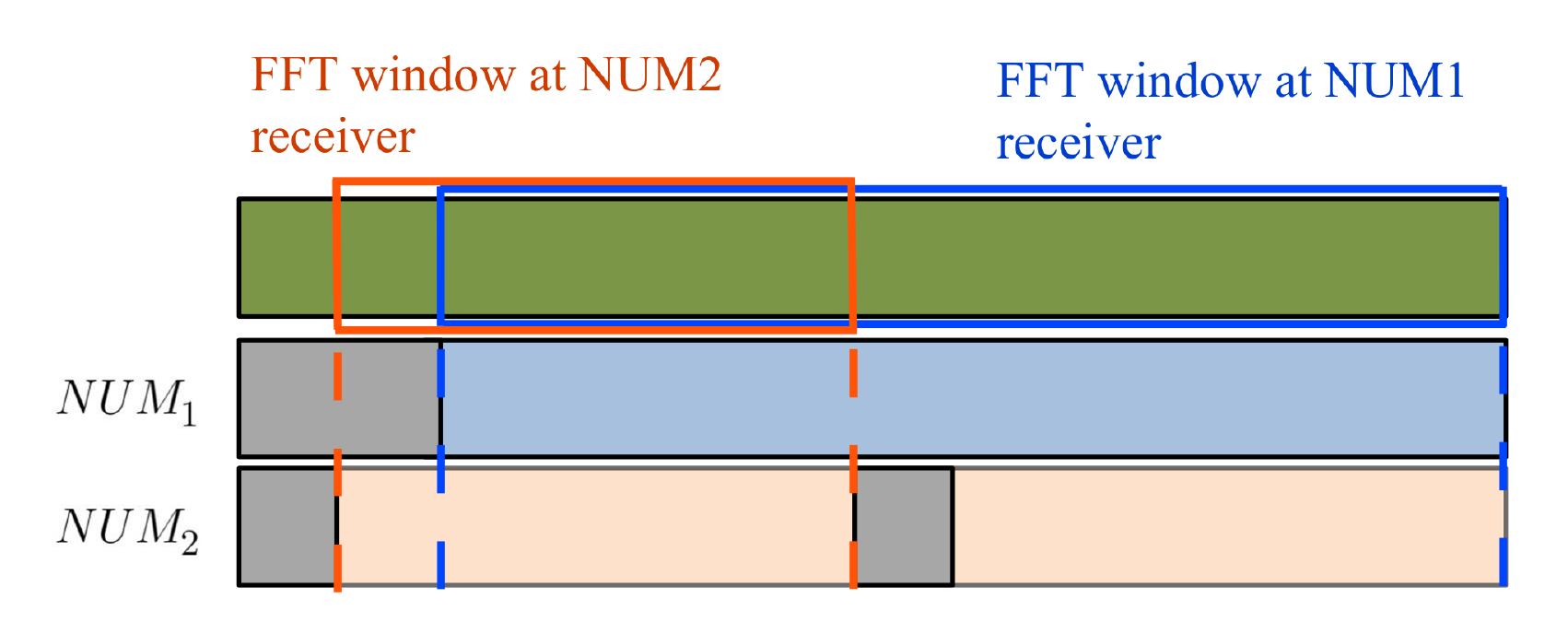} \label{fig:individualCPreceiver}
	}
	\caption{Illustration of FFT-window at the receiver of each numerology (for $k$ = 1)}
	\label{fig:FFTreceiver}
\end{figure}

\begin{figure}[h]
	\centering
	\subfigure[Common CP case]{
		\includegraphics[width=2.35in]{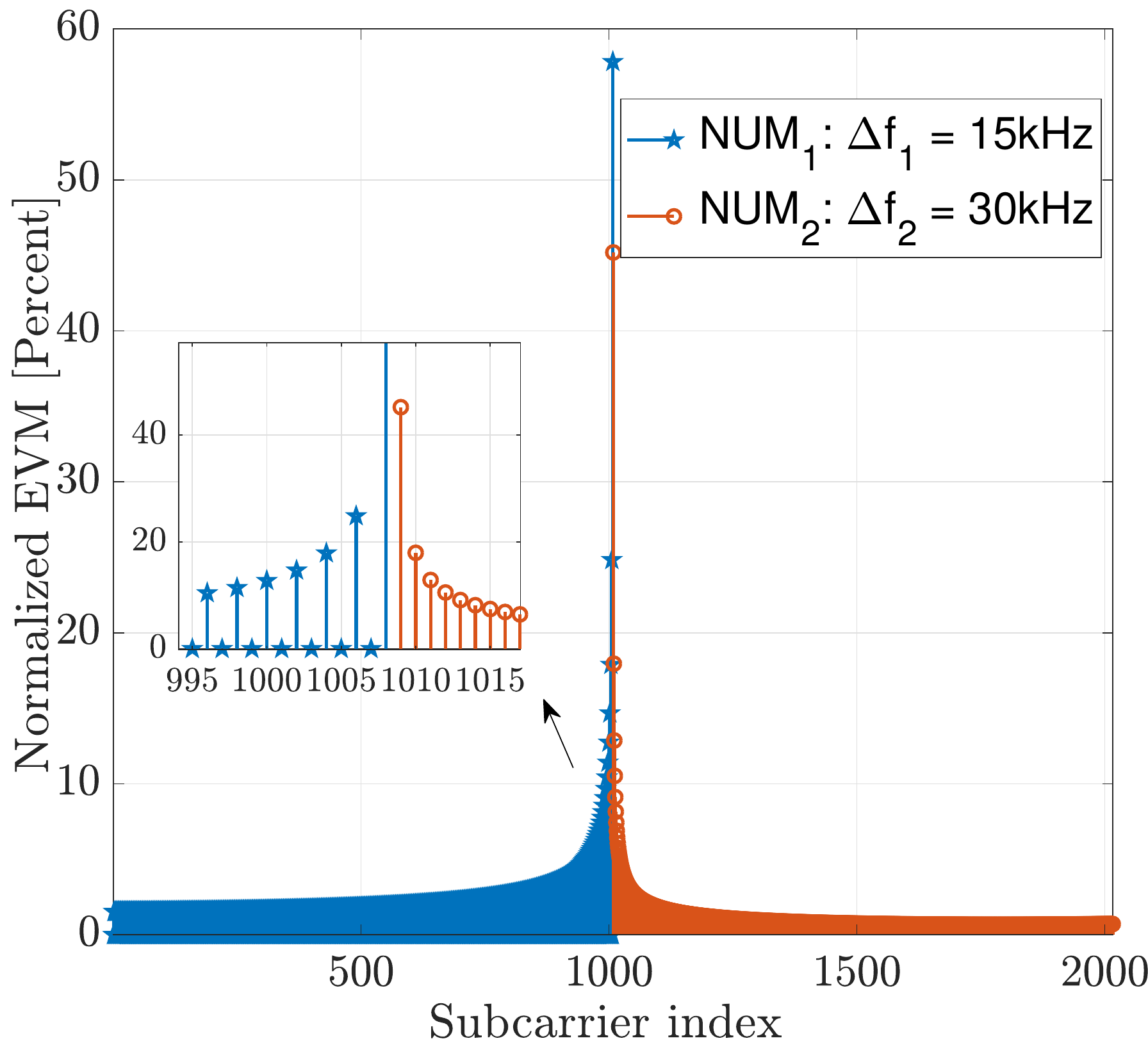} \label{fig:EVMcommonCP}
	}
	\subfigure[Individual CP case]{
		\includegraphics[width=2.35in]{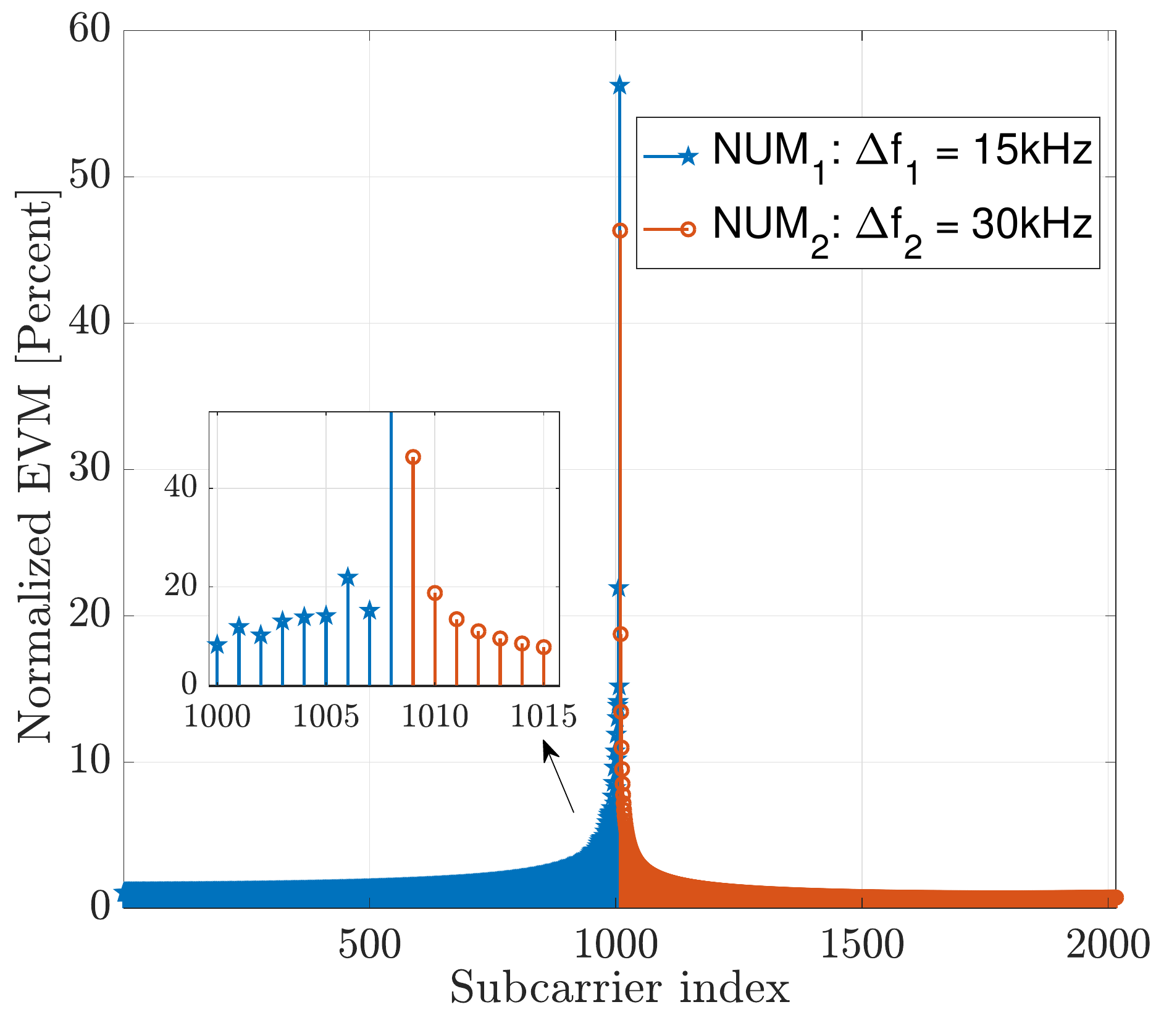} \label{fig:EVMindividualCP}
	}
	\caption{EVM plots for the two synchronization techniques}
	\label{fig:orthogonalityEVM}
\end{figure}

At the receivers, user of each numerology concentrates on capturing and decoding its own symbols from the composite signal depending on its numerology specification (Fig. \ref{fig:FFTreceiver}). \\
\textbf{\textit{Common CP case:}} Fig. \ref{fig:CommonCPreceiver} summarizes what happens at $NUM_1$ and $NUM_2$ receivers when common CP is used for synchronization. \ac{FFT} window at $NUM_1$ receiver captures a full $NUM_1$ symbol from the composite signal as well as two full symbols of $NUM_2$ as shown in Fig. \ref{fig:CommonCPreceiver} (blue window). The $N$-point \ac{FFT} (corresponding to $2*N/2^k$-point \ac{FFT} for $NUM_2$) at $NUM_1$ receiver does not disturb $NUM_2$ samples present in the composite signal (i.e $NUM_2$ subcarriers do not lose their orthogonality due to FFT process at $NUM_1$ receiver). Therefore $NUM_2$ subcarriers do not create any extra interference to $NUM_1$ at the receiver. On the other hand, when \ac{FFT} window at $NUM_2$ receiver captures one symbol of $NUM_2$ from the composite signal, it also captures a ``portion" of $NUM_1$ symbol (Fig. \ref{fig:CommonCPreceiver} (red window)). Thus, the \ac{FFT} operation at $NUM_2$ receiver causes disturbance on the $NUM_1$ samples contained in the composite signal (i.e loss of orthogonality between subcarriers of $NUM_1$ at $NUM_2$ receiver), leading to interference from $NUM_1$ to each subcarrier of $NUM_2$. The zero interferences of $NUM_1$ on the locations of each $NUM_2$ subcarrier (shown in Fig. \ref{fig:SCsTx}) will no longer be the case. Interference analysis at the receiver was done and \ac{EVM} of each subcarrier of $NUM_1$ and $NUM_2$ was observed (Fig. \ref{fig:orthogonalityEVM}). From Fig. \ref{fig:EVMcommonCP}, for common CP case, one out of every two subcarriers of $NUM_1$ has zero \ac{EVM}. This shows that interference on $NUM_1$ is the only one that was created at the transmitter (Fig. \ref{fig:SCsTx}) while all the subcarriers of $NUM_2$ are affected by \ac{INI} even though they were interference-free at the transmitter. 
 \\
\textbf{\textit{Individual CP case:}} From Fig. \ref{fig:individualCPreceiver}, we observe that \ac{FFT} window at the receiver of each numerology capture a portion of the symbol (not the full symbol) of the other numerology contained in the composite signal. Therefore \ac{FFT} process at $NUM_1$ receiver causes interference from $NUM_2$ to all subcarriers of $NUM_1$, and \ac{FFT} process at $NUM_2$ receiver causes interference from $NUM_1$ to $NUM_2$. This is revealed by Fig. \ref{fig:EVMindividualCP} where all subcarriers for each numerology are in error due to \ac{INI}.

According to the above discussion, we can say that the common CP case renders the multi-numerology system partially orthogonal while in the individual CP case, the system is totally non-orthogonal. However, the rest of simulations results presented in this study are based on the conventional individual CP case.

\section{Factors Affecting INI} \label{sec:Factors}
\subsection{Inter-Numerology Subcarrier Spacing Offset} \label{sec:SubcarrierSpacinfOffset}
Subcarrier spacing (ScS), $\Delta f$, is one of the crucial parameters in the multi-numerology concept. According to \ac{3GPP} standard document \cite{3GPP_38_211}, four options of $\Delta f$ are provided as shown in Table \ref{tab:numerology_structure}. Therefore, a numerology is free to utilize any of the standardized $\Delta f$ that suits its requirement. In this section we investigat how the choice of $\Delta f$'s among coexisting numerologies impacts the performance of multi-numerology systems.

\begin{figure}[h]
	\centering
	\subfigure[Performance of numerologies with small and large subcarriers]{
		\includegraphics[width=2.35in]{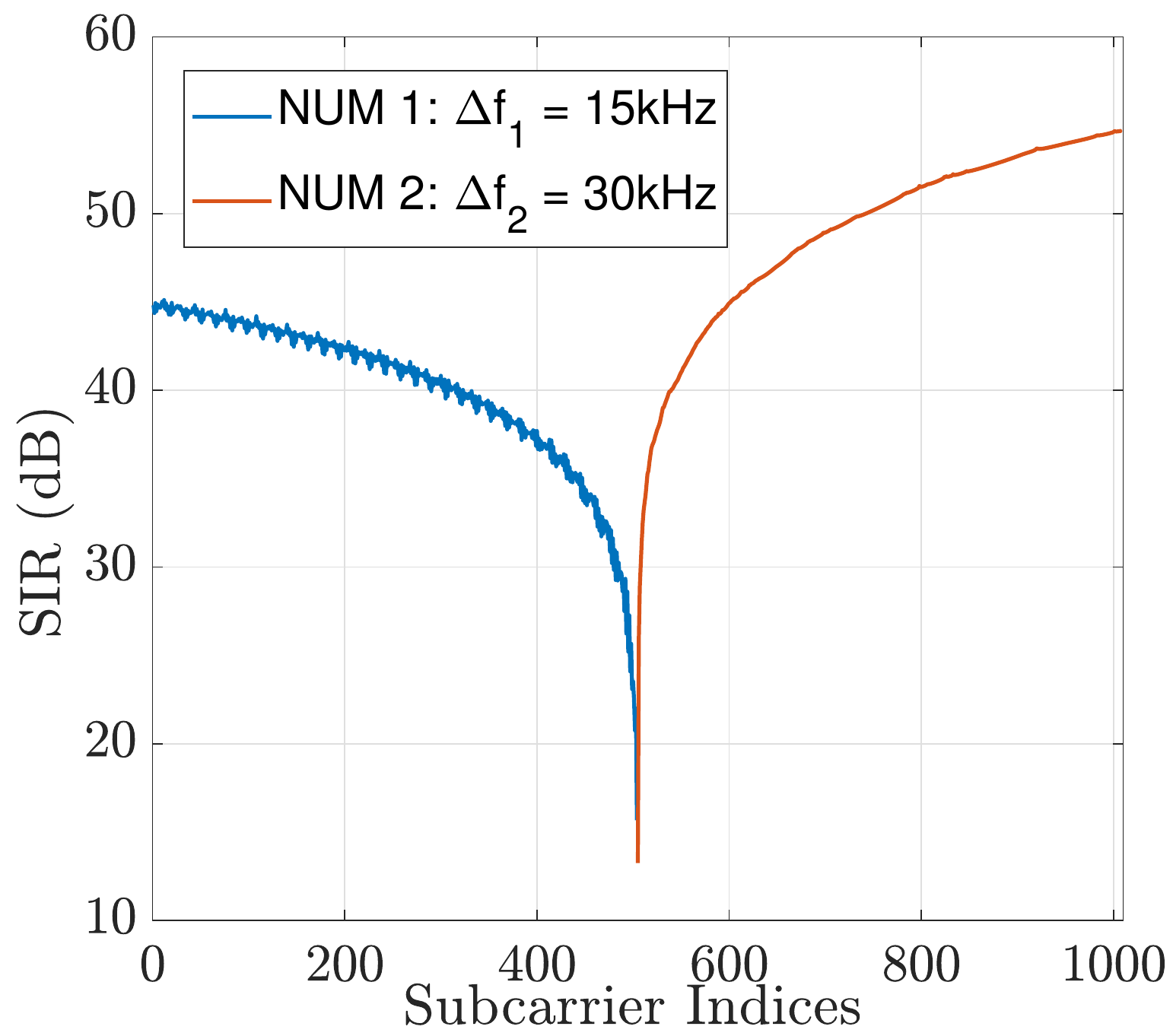} \label{fig:SCsLargeSmall}
	}
	\subfigure[Scenarios with different subcarrier spacing offsets between numerologies ]{
		\includegraphics[width=2.35in]{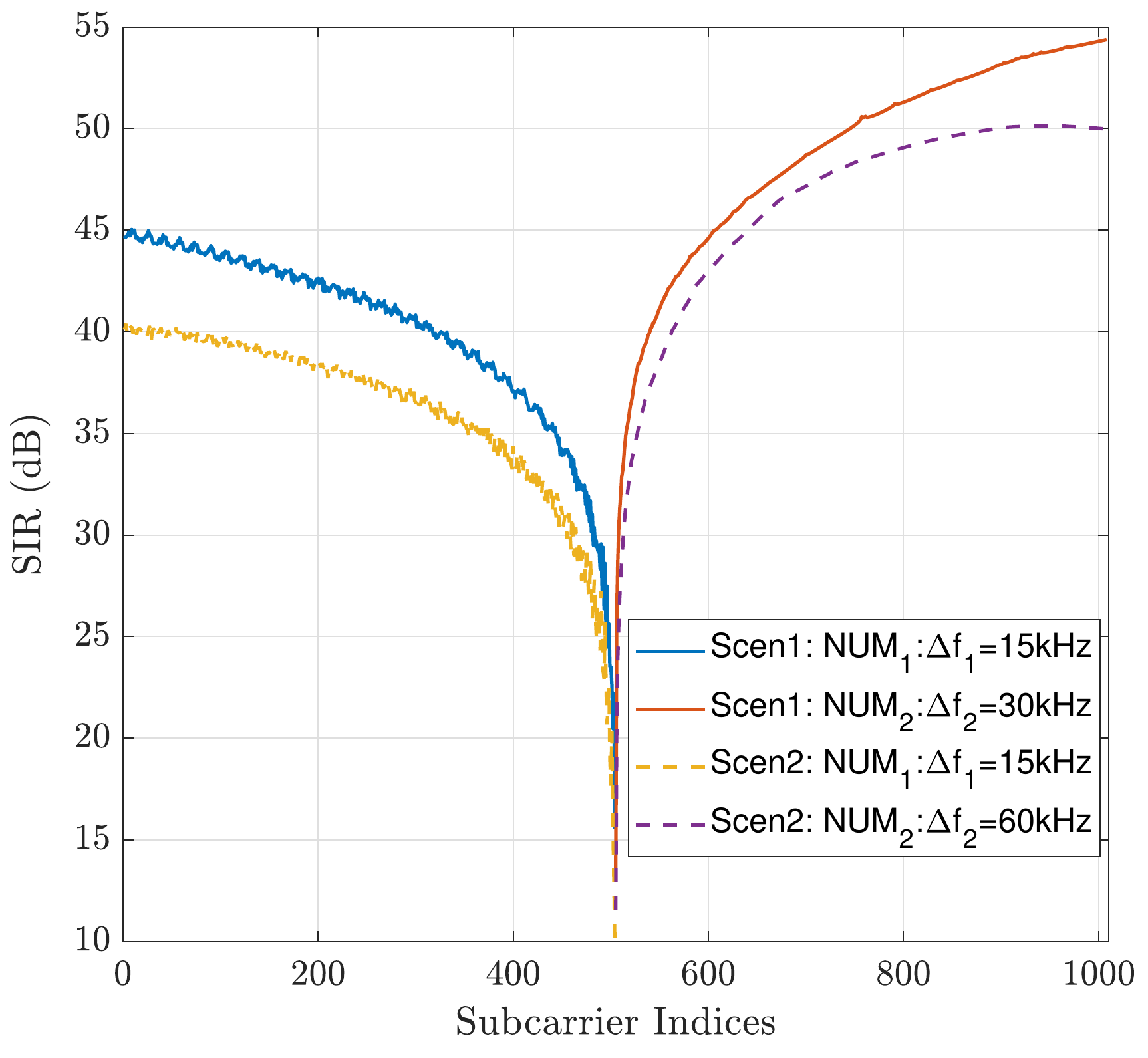} \label{fig:SCsOffset}
	}
	\caption{SIR performances of the numerologies as a function of subcarrier spacing}
	\label{fig:SubcarrierSpacingIssues}
\end{figure}

\Ac{SIR} performances of two adjacent numerologies, $NUM_1$ with $\Delta f_1$ = 15kHz, and $NUM_2$ with $\Delta f_2$ = 30kHz are observed. While all other parameters are set the same for both numerologies, $NUM_2$ exhibits better performance than $NUM_1$ (Fig. \ref{fig:SCsLargeSmall}). This result is quite expected because, as explained in the previous section (Section \ref{sec:Synchronization}), it is evident that, for individual CP case, $NUM_1$ is a victim of interference from $NUM_2$ at both, transmitter and receiver, while $NUM_2$ receives interference from $NUM_1$ only at the receiver. That's to say, numerology with small $\Delta f$ is more exposed to \ac{INI} than the one with larger $\Delta f$.

Another interesting observation regarding subcarrier spacing in the multi-numerology systems is that the \ac{SIR} performance of each numerology degrades as their \ac{SSO} (i.e $\Delta f_2$ - $\Delta f_1$) increases as shown in Fig. \ref{fig:SCsOffset}. This observation can also be linked to the discussion presented in Section \ref{sec:Synchronization}. In Fig. \ref{fig:SCsOffset}, two scenarios are presented: Scenario-1 with $\Delta f_1$/$\Delta f_2$ = 15kHz/30kHz, and Scenario-2 with  $\Delta f_1$/$\Delta f_2$ = 15kHz/60kHz.\\
\textbf{\textit{Numerology-1:}} In Scenario-1, $\Delta f_2$/$\Delta f_1$ = 2. Recalling our discussion in Section \ref{sec:Synchronization}, only one out of two subcarriers of $NUM_1$ experiences interference from $NUM_2$ at the transmitter, that is, only half of all the subcarriers of $NUM_1$ are affected by \ac{INI}. However, in Scenario-2, the ratio $\Delta f_2$/$\Delta f_1$ = 4, which causes three out of four subcarriers of $NUM_1$ to be affected by \ac{INI}. Therefore, three quarters of all subcarriers of $NUM_1$ are experiencing interference from $NUM_2$, leading to poorer \ac{SIR} performance compared to Scenario-1.\\
\textbf{\textit{Numerology-2:}} The observed degradation in $NUM_2$ can be explained from receiver side. \ac{FFT} window at the receiver of $NUM_2$ captures half of the symbol duration of $NUM_1$ from the composite signal in Scenario-1, and only a quarter of it in Scenario-2. Therefore, during \ac{FFT} operation at the receiver of $NUM_2$, Scenario-2 causes more disturbance on the samples of $NUM_1$ (and hence more severe loss of orthogonality between its subcarriers) compared to Scenario-1. This imparts higher \ac{INI} from $NUM_1$ to $NUM_2$ in Scenario-2 compared to Scenario-1. 

\subsection{Number of Subcarriers} \label{sec:Nsubcarriers}
\begin{figure}[h]
	\centering
	\includegraphics[width=3.2in]{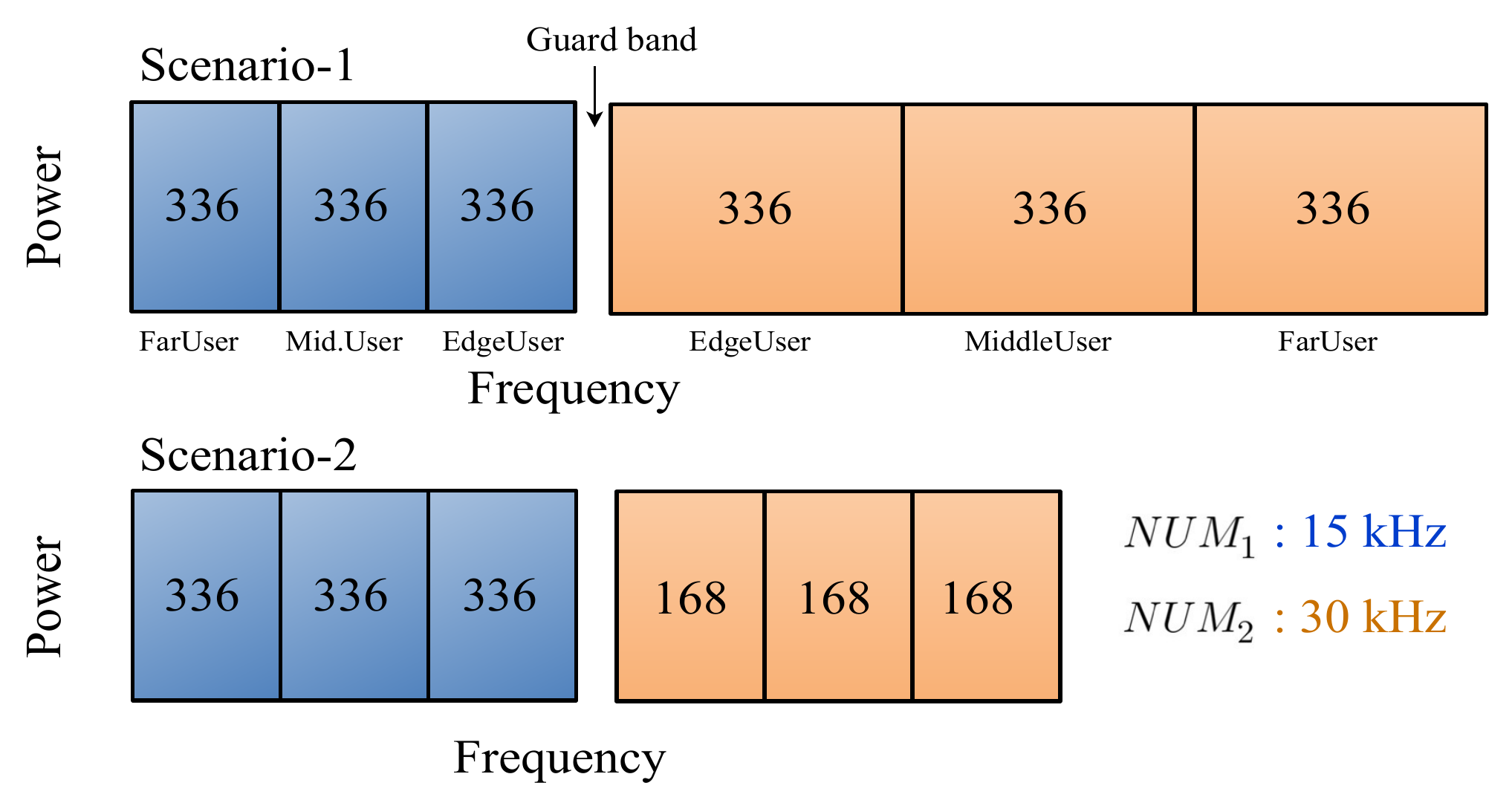}
	\caption{Scenarios for number of subcarriers}
	\label{fig:NsubcarriersScenarios}
\end{figure}

Throughput of a particular numerology can be increased by increasing its number of subcarriers. In single numerology systems larger number of subcarriers leads to the growth of \ac{PAPR} problem \cite{ochiai2001distribution}. However, in multi-numerology systems, apart from \ac{PAPR} issues, different number of subcarriers used in each numerology can be evaluated to have an impact on \ac{INI} as well. Increased number of subcarriers in one numerology corresponds to the proportional growth of its \ac{OOBE} which causes more interference to the adjacent numerology. To investigate the effect of the number of subcarriers on \ac{INI}, we considered two simple scenarios shown in Fig. \ref{fig:NsubcarriersScenarios}. Each user (in both numerologies) has 336 subcarriers in Scenario-1 and, number of Subcarriers for each user of $NUM_2$ is halved in Scenario-2.

\begin{figure}[h]
	\centering
	\includegraphics[width=3in]{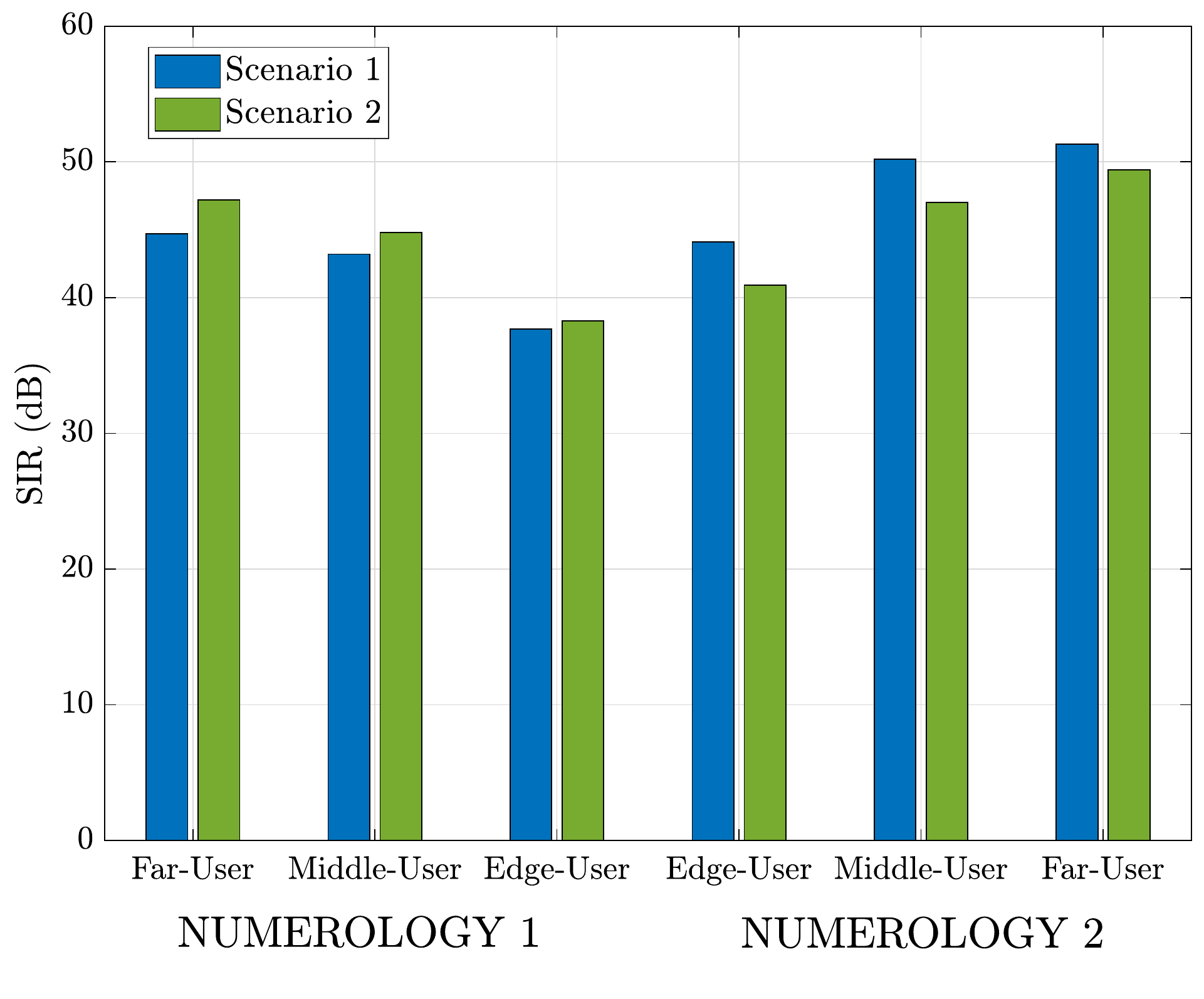}
	\caption{SIR performance for Scenario-1 and 2, with different number of subcarriers}
	\label{fig:NsubcarriersBars}
\end{figure}

Fig. \ref{fig:NsubcarriersBars} summarizes \ac{SIR} performances of the two investigated scenarios. Performances of $NUM_1$ users improve in Scenario-2 due to the less \ac{INI} they receive from $NUM_2$ as a result of the reduced number of subcarriers in $NUM_2$. Improvement in \acp{SIR} of middle and far users is higher than that of the edge user because of their larger spectral distance from $NUM_2$. The larger the distance of the user from the interfering numerology the lesser the \ac{INI} it receives. On the other hand, performance of each user of $NUM_2$ is degraded in Scenario-2. This is because, when number of subcarrier of each user is reduced, the users of $NUM_2$ get closer to $NUM_1$, exposing them to higher interference from it. 

\subsection{Power Offset} \label{sec:PowerOffset}
\begin{figure}[t]
%\begin{figure*}
    \centering
    \subfigure[No power offset between numerologies]{
	\includegraphics[width=2.35in, height = 2.0in]{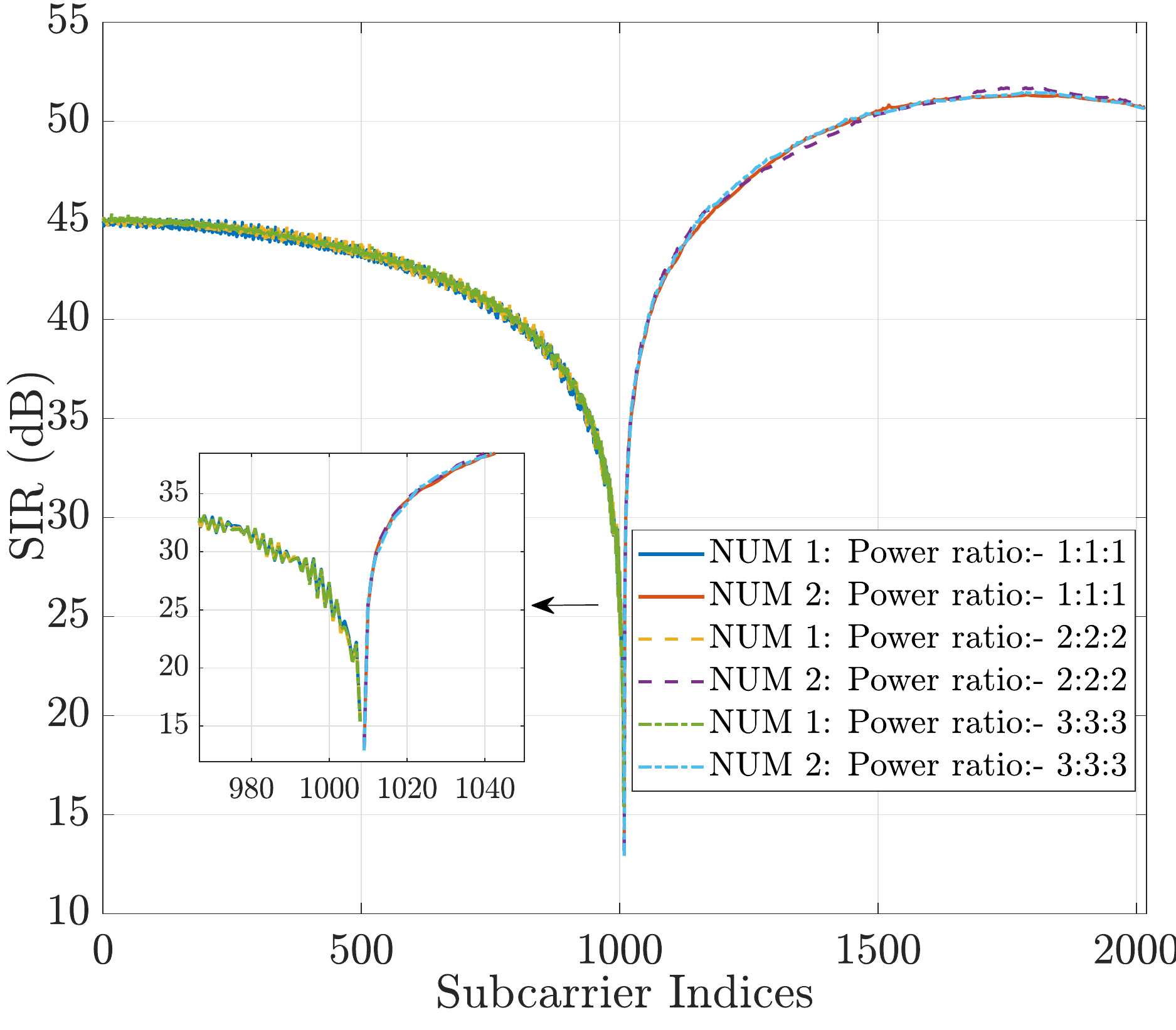}\label{fig:SamePowerRatioSIR}
}
\subfigure[With power offset between numerologies]{
	\includegraphics[width=2.35in, height = 2.0in]{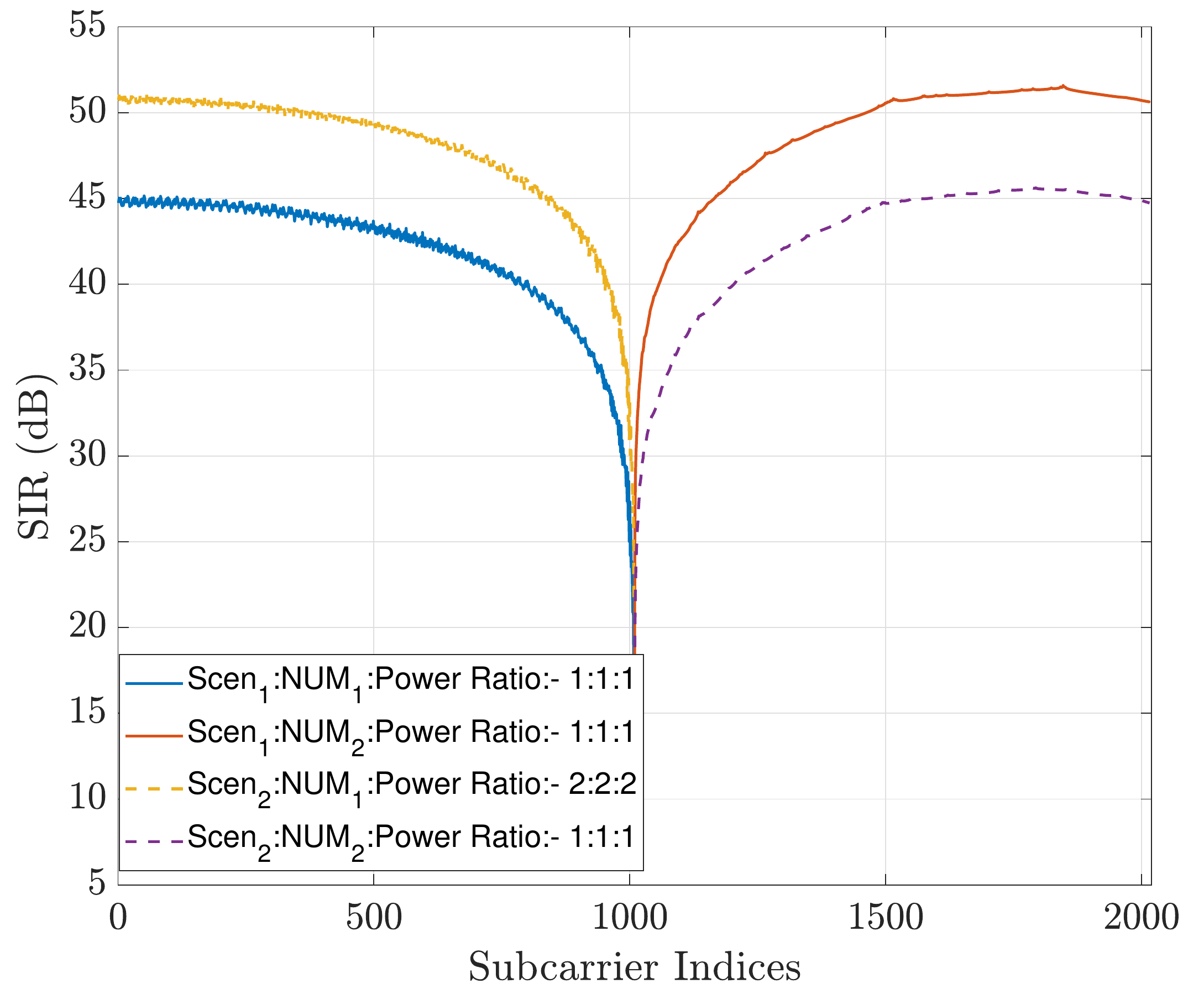}\label{fig:NumPoweroffset}
}
	\caption{SIR performance of the two numerologies as a function of their power offsets.}
	\label{fig:SIRpowerOffset}
%\end{figure*}    
\end{figure}

Users can have different power requirements depending on their channel conditions and application. Power difference among the users utilizing the same numerology does not cause any interference since the orthogonality condition is maintained. However, the \ac{$P_{off}$} between users in two adjacent numerologies significantly contributes to the amount of \ac{INI} experienced by each numerology. To illustrate this fact, let us consider an ideal case with two numerologies. In the first scenario, all users in both numerologies are assigned the same power such that \ac{$P_{off}$} in the whole system is zero. In this case the \ac{SIR} performance of each numerology remains the same regardless the actual power level assigned to the users (as long as \ac{$P_{off}$} $=$ $0$) as depicted in Fig. \ref{fig:SamePowerRatioSIR}. This is because, in this case, the amount of interference imposed on the users in the victim numerology depends solely on the spectral distance of each user from the interfering numerology. 

%% windowing figure
%\begin{figure*}[t]
\begin{figure*}[h]
	\centering
	\subfigure[Tx Windowing]{
		\includegraphics[width=2.235in]{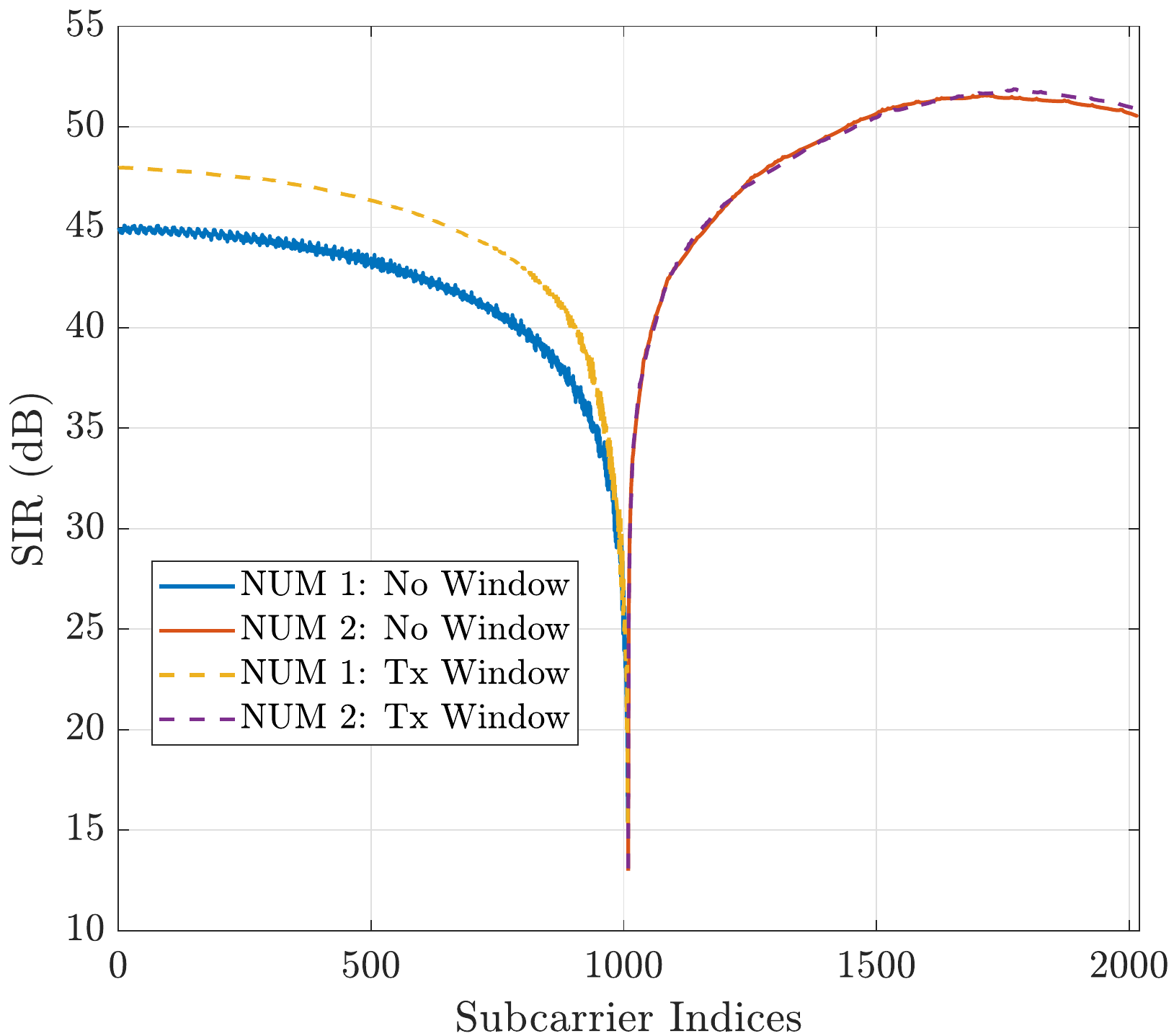} \label{fig:TxWindowing}
	}
	\subfigure[Rx Windowing]{
		\includegraphics[width=2.23in]{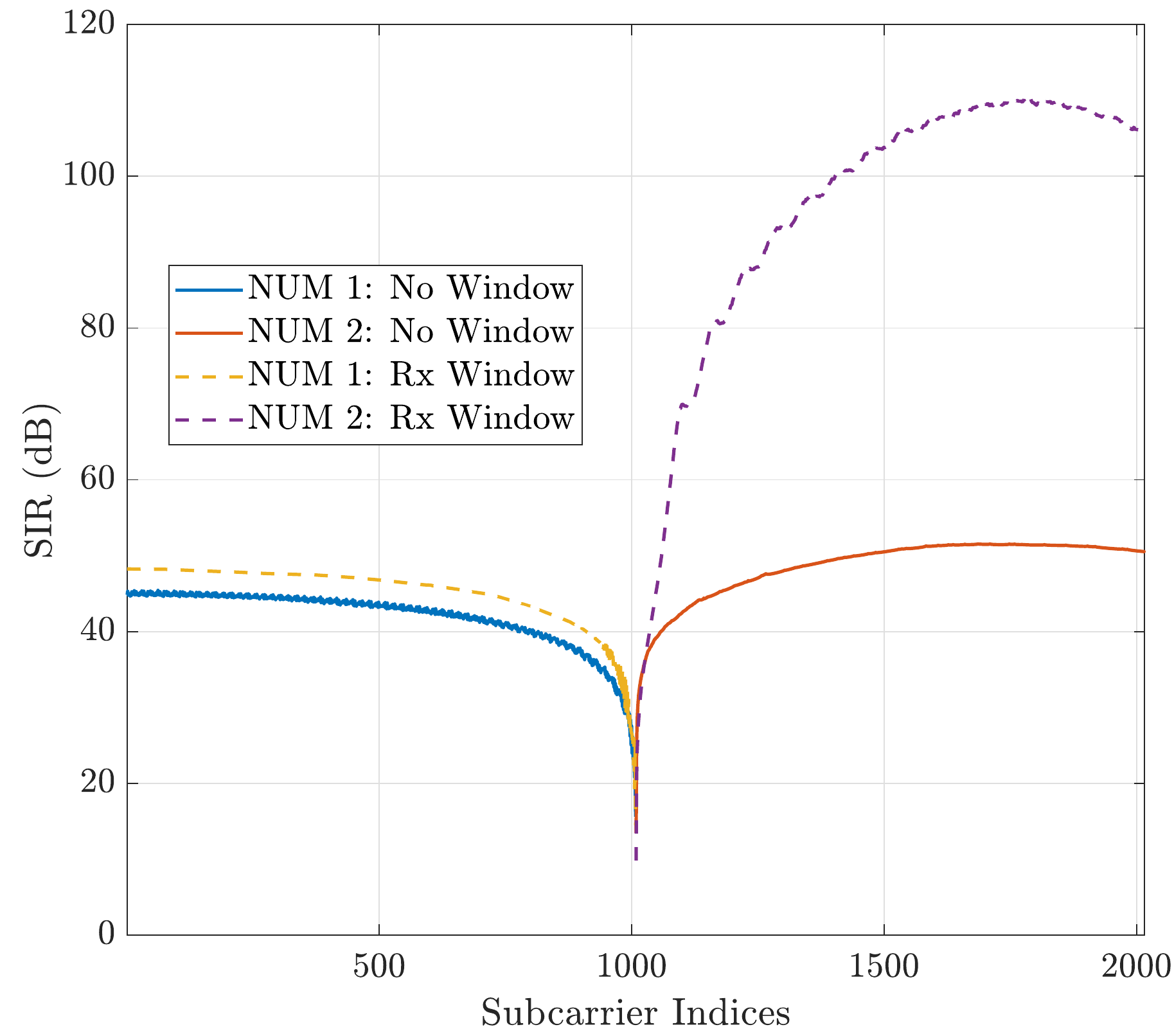} \label{fig:RxWindowing}
	}
	\subfigure[Tx Rx Windowing]{
		\includegraphics[width=2.25in]{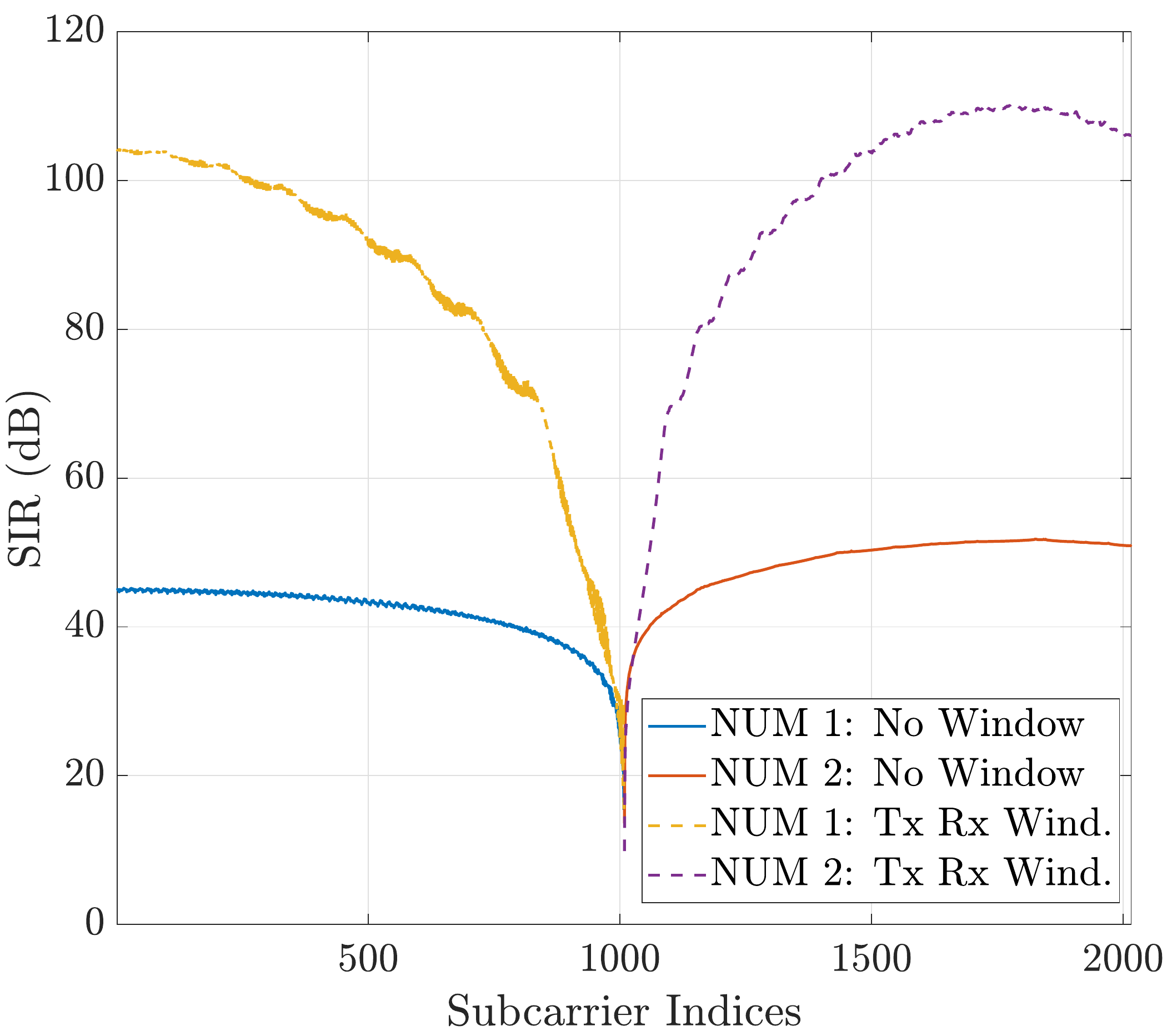} \label{fig:TxRxWindowing}
	}
	\caption{SIR performance with windowing operation (roll off factor = 0.5)}
	\label{fig:Widowing}
\end{figure*}

In the second scenario, we introduce power offset between the two numerologies. Users of the same numerology are scheduled with the same power but the power levels in the two numerologies are different. We consider the result of the first scenario (with zero \ac{$P_{off}$}) as a baseline for performance comparison. Fig. \ref{fig:NumPoweroffset} shows that, with the power offset of 3dB between the two numerologies, the performance of $NUM_2$ is degraded by about 6dB.  

The two scenarios discussed above give an idea about how critical the power offset issue can be in the performance of the multi-numerology systems. In more realistic scenarios, users are often expected to have different power requirements (even if they utilize the same numerology). Now, when power is assigned to each user according to its own need, power offset between numerologies will be random (i.e users in the victim numerology will have different power offsets with each user in the interfering numerology). In such cases, amount of \ac{INI} experienced by each user in the victim numerology depends not only on the spectral distance of that user from the interfering numerology but also its power offset with each user of the interfering numerology. Proper scheduling technique would be required to minimize the power offsets and optimize performance of each user in the multi-numerology systems~\cite{arslan2018fairness}. 

\subsection{Windowing} \label{sec:Windowing}
High out of band emission of the \ac{OFDM} waveform can be reduced by smoothing the edges of its rectangular pulse. One way of achieving this is through a windowing process where each \ac{OFDM} symbol is multiplied with a smooth function, such as raised cosine. Windowing can be applied either at the transmitter or receiver, or both. Transmitter windowing reduces the possible spectral leakage to the adjacent numerology whereas receiver windowing provides a better interference rejection at the receiver \cite{bala2013shaping}. Now, with reference to our discussion in Section \ref{sec:Synchronization}, for the two numerologies $NUM_1$, and $NUM_2$ shown in Fig. \ref{fig:SCsTx}, $NUM_1$ receives \ac{INI} from $NUM_2$ at both, transmitter and receiver. Therefore, applying both, transmitter windowing on $NUM_2$ and receiver windowing on $NUM_1$ should significantly improve \ac{SIR} performance of $NUM_1$. Also, $NUM_2$ receives \ac{INI} from $NUM_1$ only at the receiver. Therefore, receiver windowing on $NUM_2$ is expected to be sufficient enough to enhance \ac{SIR} performance of $NUM_2$. This agrees well with the simulation results presented in Fig. \ref{fig:Widowing}.

Simulation was conducted for $NUM_1$ with $\Delta f_1$ = 15 kHz and $NUM_2$ with $\Delta f_2$ = 30 kHz, and the raised cosine window was employed by adopting the steps discussed in \cite{bala2013shaping}. Fig. \ref{fig:TxWindowing} shows that transmitter windowing on $NUM_1$ does not cause any significant improvement on performance of $NUM_2$ since power leakage from $NUM_1$ to the subcarriers of $NUM_2$ at the transmitter is already zero (see Fig. \ref{fig:SCsTx}). However, transmitter windowing on $NUM_2$ enhances the SIR performance of $NUM_1$ to some extent. Fig. \ref{fig:RxWindowing} reveals the effect of receiver windowing when applied alone. Again, receiver windowing on $NUM_1$ only slightly improves its performance, while outstanding performance is achieved on $NUM_2$ with receiver windowing for the same roll off factor. Finally, Fig. \ref{fig:TxRxWindowing} shows that combination of transmitter and receiver windowing significantly improves performance of $NUM_1$ compared to the case when they are applied alone. For $NUM_2$, the \ac{SIR} performance with transmitter and receiver windowing is quite the same with the case when receiver windowing is applied alone. In summary, better \ac{SIR} performance of the numerology with small \ac{ScS} can be achieved with both, transmitter windowing on the interfering numerology as well as receiver windowing at its own receiver. But for numerology with larger \ac{ScS}, only receiver windowing can be enough.

\begin{figure}
	\centering
	\includegraphics[width=3.5in]{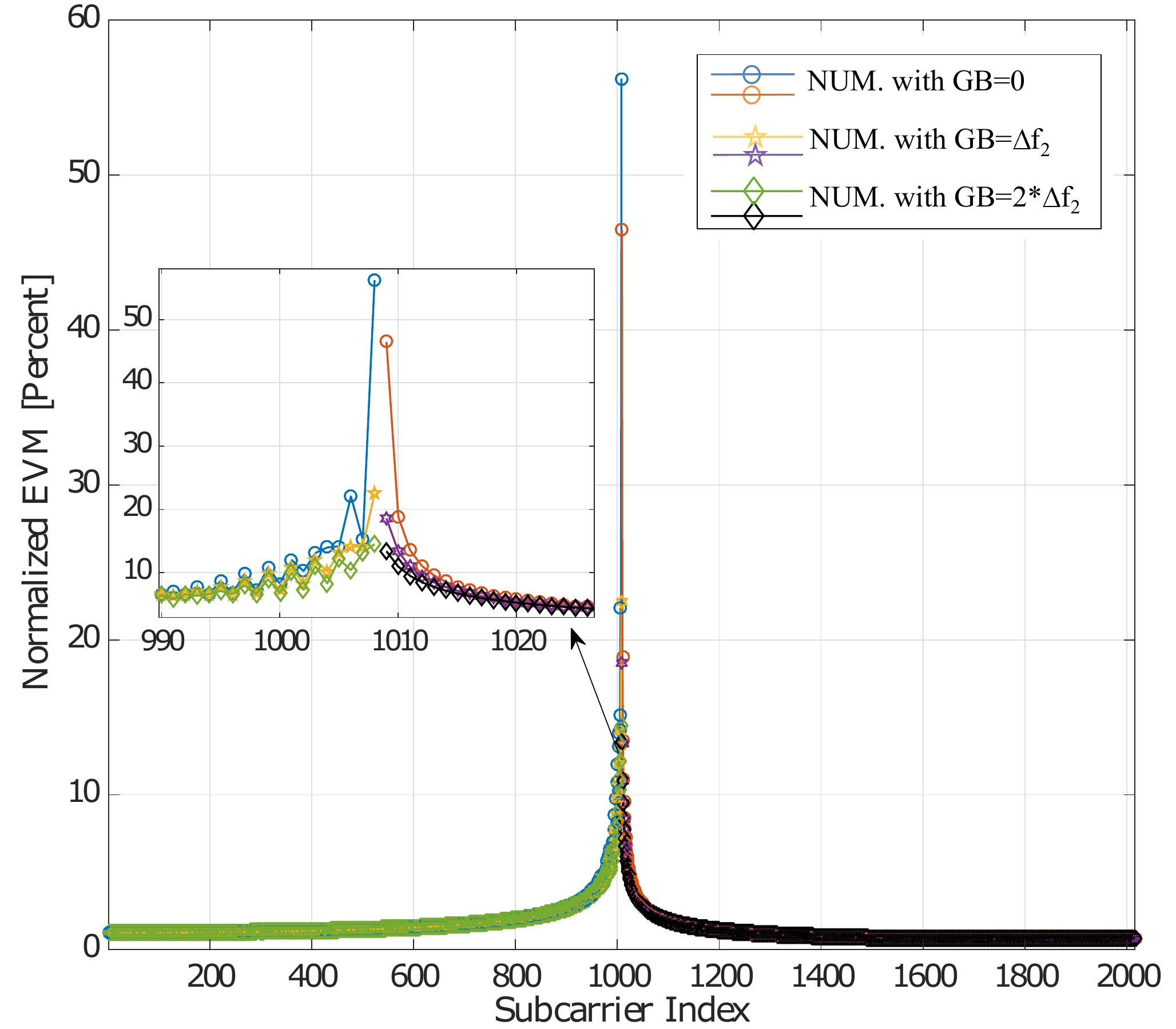}
	\caption{EVM plot for different amounts of guard band between numerologies}
	\label{fig:GuardbandEVM}
\end{figure}
\subsection{Guard Band} \label{sec:Guardband}
 Employing \acp{GB} between adjacent numerologies is another way of reducing the effect of \ac{INI} at the expense of spectral efficiency of the system. Our simulation result for two numerologies $NUM_1$ with $\Delta f_1$ = 15kHz, and $NUM_2$ with $\Delta f_2$ = 30kHz shows that \ac{GB} is effective in improving \ac{SIR} performance of the edge subcarriers only. No significant improvement is observed for subcarriers far from the edges as shown in Fig. \ref{fig:GuardbandEVM}.

\section{Conclusion}
\label{sec:Conclusion}
Next generations of wireless systems are geared towards ultimate flexibility in different aspects. An introduction of multi-numerology concept as a part of this flexibility has brought new problems, such as \ac{INI}, that require special attention from researchers. This paper has investigated the \ac{INI} problem and intuitively explained its underlying causes from signal processing point of view at the transmitter and receiver. The paper goes further and investigates the performance of multi-numerology system when coexisting numerologies flexibly adopt different parameters such as subcarrier spacing, number of subcarriers, power, etc. All the relationships observed in this study are supported by simulation results, however it will be extended to provide a thorough mathematical analysis of the \ac{INI} and the factors affecting it.

%\appendices
%\section{appendix one}\label{sec:Appendix1}
%appendix one
%	
%\section{appendix two}\label{sec:Appendix2}
%appendix two

% biography section
\bibliography{IEEEfull,GlobecomReferences}
\bibliographystyle{IEEEtran}
\end{document}

%% file: AcronymList.tex
\acsetup{list-long-format=\capitalisewords}

\DeclareAcronym{AOA}{short= AOA, long= angle of arrival}
\DeclareAcronym{AOD}{short= AOD, long= angle of departure}
\DeclareAcronym{CDF}{short= CDF, long= cumulative distribution function}
\DeclareAcronym{CSI}{short= CSI, long= channel state information}
\DeclareAcronym{comp}{short= CoMP, long= coordinated multipoint}
\DeclareAcronym{DL}{short= DL, long= downlink}
\DeclareAcronym{EM}{short= EM, long=  electromagnetic}
\DeclareAcronym{FCC}{short= FCC, long= Federal Communications Commission}
\DeclareAcronym{FoV}{short= FoV, long= field of view}
\DeclareAcronym{HO}{short= HO, long= handover}
\DeclareAcronym{LOS}{short= LOS, long= line-of-sight}
\DeclareAcronym{LTE}{short= LTE, long= long-term evolution}
\DeclareAcronym{mmwave}{short= mmWave, long= millimeter-wave}
\DeclareAcronym{MPC}{short= MPC, long= multipath component}
\DeclareAcronym{NLOS}{short= NLOS, long= non-line-of-sight}
\DeclareAcronym{OFDM}{short= OFDM, long= Orthogonal Frequency Division Multiplexing}
\DeclareAcronym{OFDMA}{short= OFDMA, long= Orthogonal Frequency Division Multiple Access}
\DeclareAcronym{PDF}{short= PDF, long=  probability density function}
\DeclareAcronym{PDP}{short= PDP, long=  power delay profile}
\DeclareAcronym{PSD}{short= PSD, long=  power spectral density}
\DeclareAcronym{RF}{short= RF, long=  radio frequency}
\DeclareAcronym{RMS}{short= RMS, long= root mean square}
\DeclareAcronym{RSRP}{short= RSRP, long= reference signal received power}
\DeclareAcronym{RVC}{short= RVC, long= Reverberation chamber}
\DeclareAcronym{SAW}{short= SAW, long=  surface acoustic wave}
\DeclareAcronym{SINR}{short= SINR, long= signal to interference plus noise ratio}
\DeclareAcronym{SON}{short= SON, long= self optimizing network}
\DeclareAcronym{TP}{short= TP, long= transmission point}
\DeclareAcronym{UL}{short= UL, long= uplink}
\DeclareAcronym{UE}{short= UE, long= user equipment}
\DeclareAcronym{IDT}{short= IDT, long= interdigital transducer}
\DeclareAcronym{CIR}{short= CIR, long= channel impulse response}
\DeclareAcronym{$LCR_f$}{short= $LCR_f$, long= level crossing rate}
\DeclareAcronym{ABF}{short= ABF, long= average bandwidth of fade}
\DeclareAcronym{INI}{short= INI, long= inter-numerology interference}
\DeclareAcronym{SIR}{short= SIR, long= Signal to Interference Ratio}
\DeclareAcronym{$P_{off}$}{short= $P_{off}$, long= power offset}
\DeclareAcronym{ScS}{short= ScS, long= subcarrier spacing}
\DeclareAcronym{LCM}{short= LCM, long= least common multiplier}
\DeclareAcronym{MSE-OFDM}{short= MSE-OFDM, long= multi symbols encapsulated OFDM}
\DeclareAcronym{FFT}{short= FFT, long= Fast Fourier Transform}
\DeclareAcronym{EVM}{short= EVM, long= error vector magnitude}
\DeclareAcronym{SSO}{short= SSO, long= subcarrier spacing offset}
\DeclareAcronym{PAPR}{short= PAPR, long= peak-to-average power ratio}
\DeclareAcronym{OOBE}{short= OOBE, long= out of band emission}
\DeclareAcronym{3GPP}{short= 3GPP, long= 3rd Generation Partnership Project}
\DeclareAcronym{GB}{short= GB, long= guard band}